\documentclass[3p,times]{elsarticle}

\usepackage{ecrc}
\newcommand{\sqrts}{$\sqrt{s}$}
\newcommand{\pT}{$p_{T}$}

\volume{00}

\firstpage{1}

\journalname{Nuclear Physics A}

\runauth{}

\jid{nupha}

\usepackage{amssymb}

\usepackage[figuresright]{rotating}

\begin{document}

\begin{frontmatter}



\dochead{}

\title{Jet and Underlying Event Measurements in p+p collisions at RHIC }


\author{Helen Caines for the STAR Collaboration}

\address{Physics Department, Yale University, New Haven, CT 06520, U.S.A.}

\begin{abstract}

The physics of hadron-hadron collisions is very complex involving both 
perturbative and non-perturbative QCD. It is therefore imperative to 
study p-p collisions in as much detail as possible to provide a wide 
variety of data against which the various theoretical calculations can 
be tested. Direct jet measurements, for instance,  help address 
fundamental questions of the fragmentation process. These measurements 
form a critical baseline for comparisons of results from heavy-ion 
studies, where modifications of the fragmentation functions are expected 
due to interactions of the high ${Q^2}$ scattered partons with the hot and 
dense medium. Finally, it is also important to gain a deeper 
understanding of how the beam-beam remnants, multi-parton interactions 
and initial- and final-state radiation combine to produce the particles 
observed in the underlying event. In this talk we present results on jet 
production and the underlying event in p-p collisions at 200 GeV 
collisions as measured by the STAR experiment at RHIC.

\end{abstract}

\begin{keyword}
RHIC jet p-p fragmentation

\end{keyword}

\end{frontmatter}

\section{Introduction, the Dataset and the Analysis}

Our understanding of QCD and the hadronization process in particular can be improved by studying jet production and the properties of the underlying event  in p-p collisions. In addition these results serve as a baseline for measurements being performed in heavy-ion collisions~\cite{JetTalks}. Preliminary data 
from  \sqrts  = 200 GeV collisions at RHIC are presented.  Both minimum bias and ``jet-patch" triggers were recorded;  a jet patch is defined as a $\Delta \eta \times\Delta \phi$ = 1$\times$1 patch of the Barrel Electromagnetic Calorimeter (BEMC) containing transverse energy, E$_{T}>$8 GeV. Such a trigger  creates a neutral energy fragmentation bias for the triggering jet, hence fragmentation functions are presented only  for the di-jet partner. Jets were reconstructed using the FastJet  package's
 k$_{T}$ and anti-k$_{T}$ recombination and the  SISCone jet algorithms~\cite{FastJet} from charged and neutral particles measured in the Time Projection Chamber, TPC,  and BEMC respectively.
 A cut of \pT\ or $E_{T}$  $>$0.2 GeV/c (for charged and neutral hadrons respectively) was applied to all particles considered in the event.  Jet resolution parameters of R=0.4 or 0.7 were used for the analyses reported. It has been shown that  a resolution parameter of R=0.4 encompasses more the 75$\%$ of the jet's energy in p-p collisions at RHIC~\cite{Caines:2009iy}. 
 
\begin{figure}[htb]
		\begin{center}
			\includegraphics[width=0.6\linewidth]{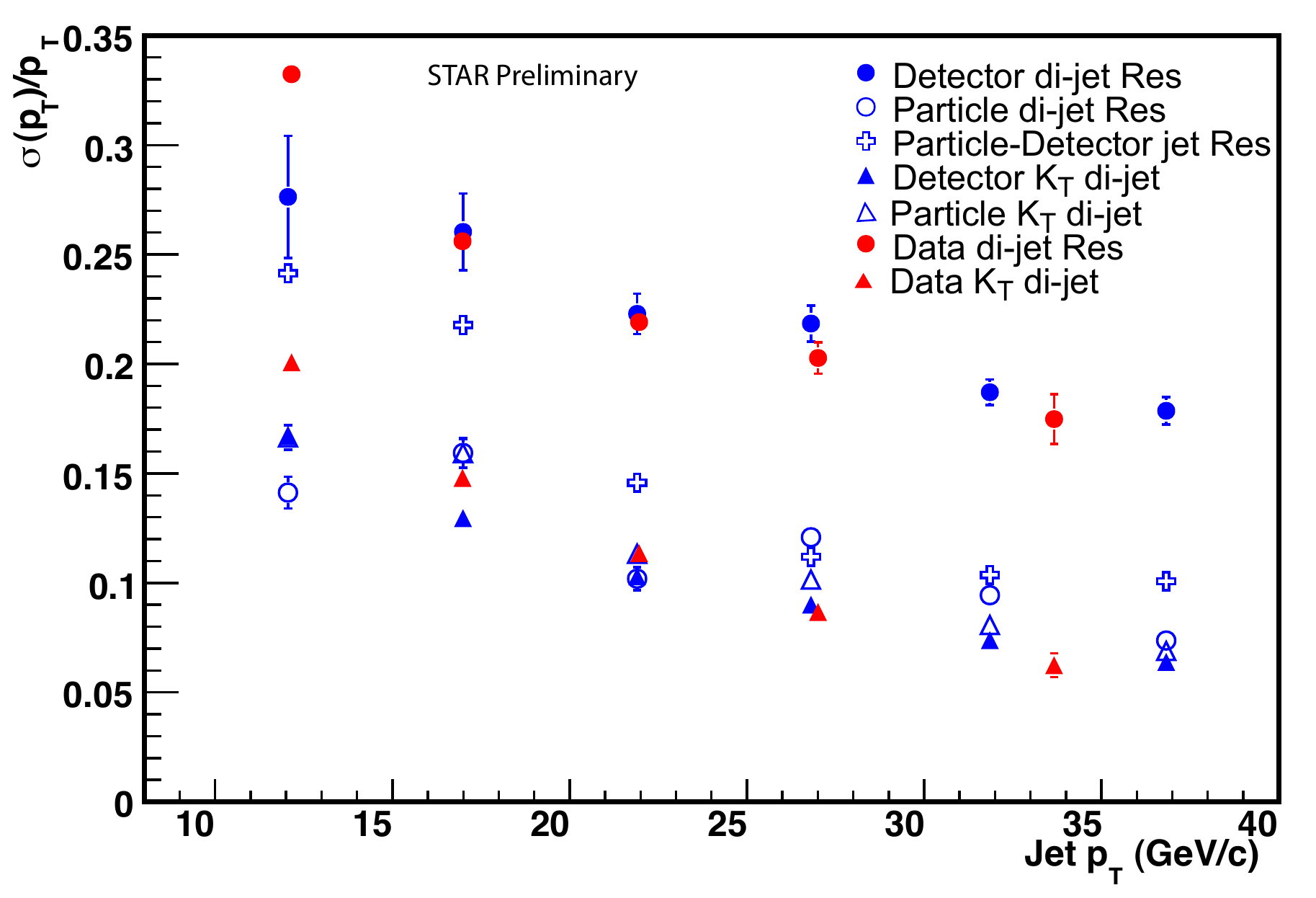}
		\end{center}
	\caption{ Color online: $\sigma(p_{T})/p_{T}$ from simulations and p-p data as a function of reconstructed jet \pT\ for anti-k$_{T}$, R=0.4. Particle =  Pure hadronic level simulation. Detector = Particle level simulation + detector effects such as resolutions and inefficiencies. Particle-Detector jet Res = Difference in relative jet energy  resolutions as calculated at the Particle and Detector levels. See text for further details.}
	\label{Fig:JetRes}
\end{figure}

Before  interpreting the jet measurements, it is vital that STAR's jet energy  resolution be determined. To this end PYTHIA 6.410~\cite{Sjostrand:2006za} events, tuned to the CDF 1.96 TeV data (Tune A), and passed through STAR's simulation and reconstruction algorithms have been studied.  Jets reconstructed at the particle level are compared to jets reconstructed at the detector level, i.e. including all the detectors'  irresolutions and inefficiencies.  The relative jet energy resolution, $\sigma(p_{T})/p_{T}$, is determined from the width of the difference in matched jet energies at the detector and particle level. As can be seen, the open blue crosses in Fig.~\ref{Fig:JetRes}, for R=0.4 the resolution varies from 10-25$\%$ depending on the  jet \pT. To ensure that the simulation matches the data we proceeded with further studies. First the k$_{T}$ of the jets was determined from the width of the di-jet E$_{jet}$sin$\Delta\phi_{jet}$ distribution, where $\Delta\phi_{jet}$ is the azimuthal angle between the di-jet pair. Reasonable agreement is seen in Fig.~\ref{Fig:JetRes}  between the data, particle and detector level measurements. For more details on the k$_{T}$ measurement see ~\cite{JanKt}. Next the di-jet energy balance was measured and the width of the resulting distribution studied as a function of jet \pT\ for data and detector and particle level simulations; the closed red and blue circles and open blue circles in Fig.~\ref{Fig:JetRes} respectively.  This measure is a folding of  both the jet energy resolution and k$_{T}$ smearing. The  width of the particle level di-jet energy difference matches the particle level k$_{T}$ measurements. This is expected since  the only cause of a di-jet energy imbalance at the particle level is the k$_{T}$. Good agreement is observed between the di-jet resolution for the real data and detector simulations. This measure is significantly higher than both the true resolution and the measured k$_{T}$.  The combination of results presented in Fig.~\ref{Fig:JetRes} indicate that the detectors are well understood and simulated,  hence the resolution determined from the PYTHIA simulations can be successfully applied to the p-p data.

\section{Results}

\begin{figure}[htb]
	\begin{minipage}{0.46\linewidth}
		\begin{center}
			\includegraphics[width=\linewidth,]{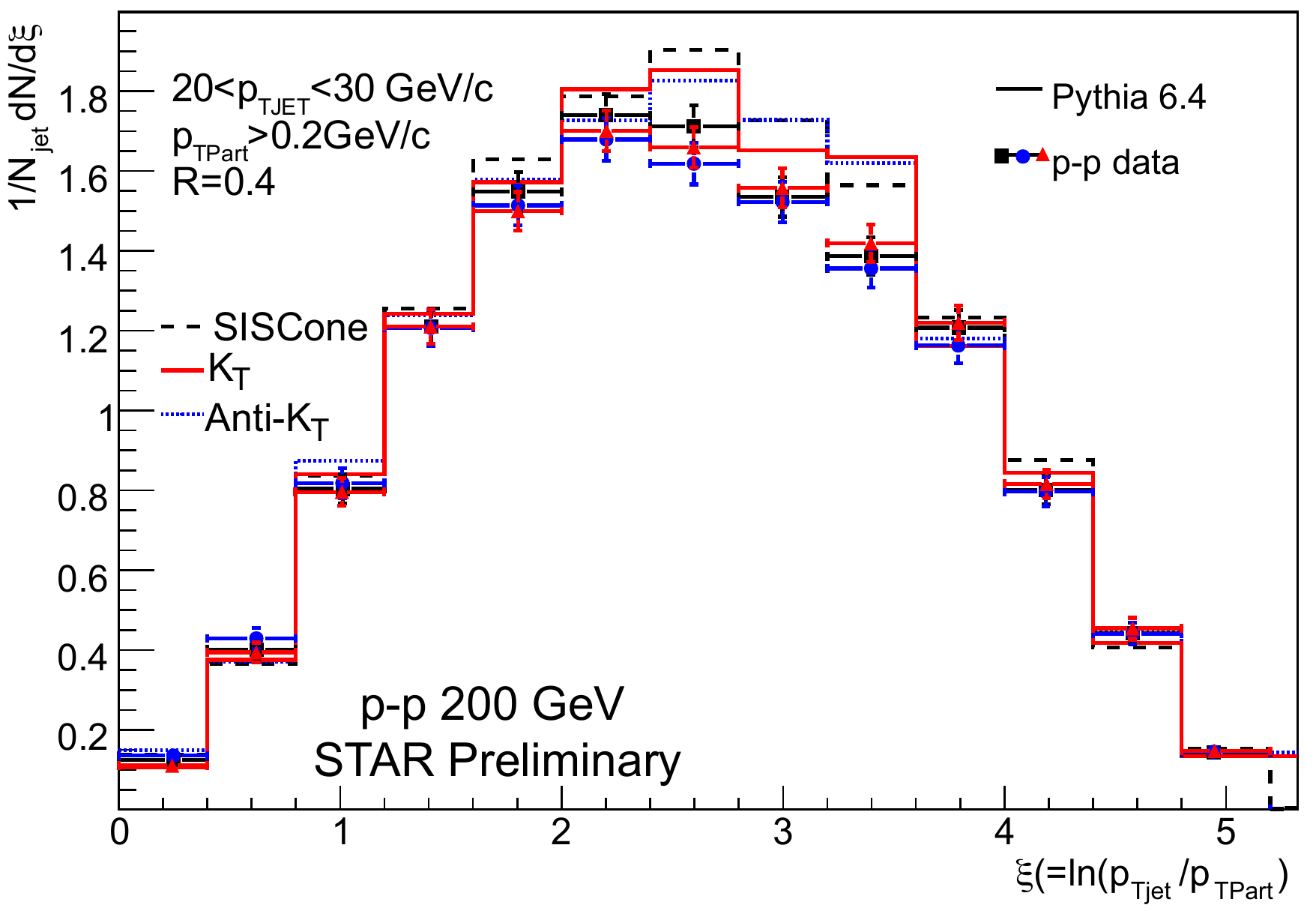}
		\end{center}
	\end{minipage}
	\begin{minipage}{0.46\linewidth}
		\begin{center}
		\includegraphics[width=\linewidth]{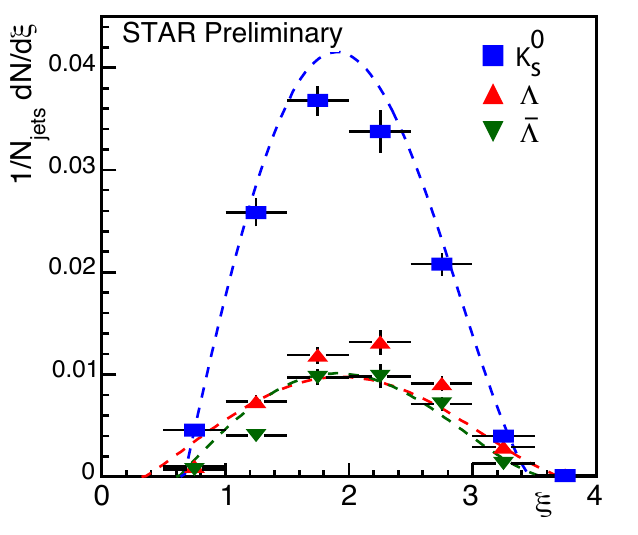}
		\end{center}
	\end{minipage}
	\caption{ Color online: Detector level FF as a function of $\xi$ for R=0.4 $|\eta|<$1-R  Left) Charged particles  for reconstructed jets  with 20$<$ \pT$<$ 30 GeV/c, compared to PYTHIA (solid histograms) for 3 different jet algorithms. Right) Strange particles for reconstructed jets  with 20$<$ \pT$<$ 40 GeV/c. The curves  show polynomial fits to the PYTHIA  predictions.}
			\label{Fig:FF04}
\end{figure}

The inclusive jet spectrum~\cite{Abelev:2006uq} and minimum bias $\pi$ and proton spectra~\cite{Adams:2006nd} have been measured by STAR and are in good agreement with NLO pQCD calculations over several orders of magnitude. In addition preliminary charged particle fragmentation functions, FF,  have been measured.  These data are not yet corrected to the particle level so comparisons are instead made to the detector level PYTHIA simulations discussed above. Figure~\ref{Fig:FF04} (left panel) shows the results for 20-30 GeV jets for the three jet algorithms, the histograms are the PYTHIA predictions. It can be seen that there is reasonable agreement between the data and PYTHIA. A similar agreement is observed for R=0.7~\cite{Caines:2009iy}. This similarity, especially for the larger resolution parameter, suggests that there are only minor NLO contributions beyond those mimicked in the PYTHIA parton-shower calculations at RHIC energies. 

While good agreement between the data, NLO calculations, and PYTHIA predictions is seen for inclusive charged particles and  for $\pi$ and protons, theoretical descriptions  for strange particle production (K, $\Lambda$ and $\Xi$) significantly under predict the measured yields at intermediate to high \pT\ \cite{Abelev:2006cs}. If the K factor is increased in PYTHIA from K=1 to K=3,  reasonable agreement to the strange baryon data can be obtained, however,  the  $\pi$ and proton data are then no longer reproduced.  Strange hadron production in jets is therefore being studied to further investigate strangeness production in p-p collisions.  K$^0_S$, $\Lambda$, and $\bar{\Lambda}$ are reconstructed via their V$^{0}$ decay topologies. Good signal-to-noise ratios are obtained, after applying geometrical cuts, for those particles with \pT $>$ 1 GeV/c. The small residual backgrounds under the mass peaks are not yet corrected for.  Figure~\ref{Fig:FF04} (right panel) shows the measured detector level FF; again the curves are from PYTHIA.  The uncertainties are statistical added in quadrature to the small differences obtained by comparing the three different jet finders.  PYTHIA gives an reasonable description of the $K^{0}_{S}$ data over the measured range. This is in agreement with  $e^{+}+e^{-}$ collision $K^{\pm}$ FF, for  $5 < E_{jet} < 46$ GeV,   where JETSET calculations (the jet production scheme in PYTHIA) also described the data~\cite{Albino:2008afa}.  Although the strange baryon  predictions from PYTHIA have the correct over-all yields in the range measured, the trend as a function of $\xi$ is incorrect, over predicting at low $\xi$ and under predicting  at intermediate $\xi$. The integrated $\bar{\Lambda}/\Lambda$ and $\Lambda/K^{0}_{S}$ ratios for $p_{T} > 1$ GeV/c as a function of reconstructed jet $p_T$~are consistent with the values obtained from the minimum bias inclusive spectra when the same \pT\ range is considered~\cite{Timmins:2010ey}. These results suggest that the  $\Lambda$ and $K^{0}_{S}$ spectra for $p_{T} > 1$ GeV/c have a dominant contribution from hard processes, i.e. jet production. Further studies are needed to confirm this.

\begin{figure}[htb]
	\begin{minipage}{0.46\linewidth}
		\begin{center}
			\includegraphics[width=\linewidth]{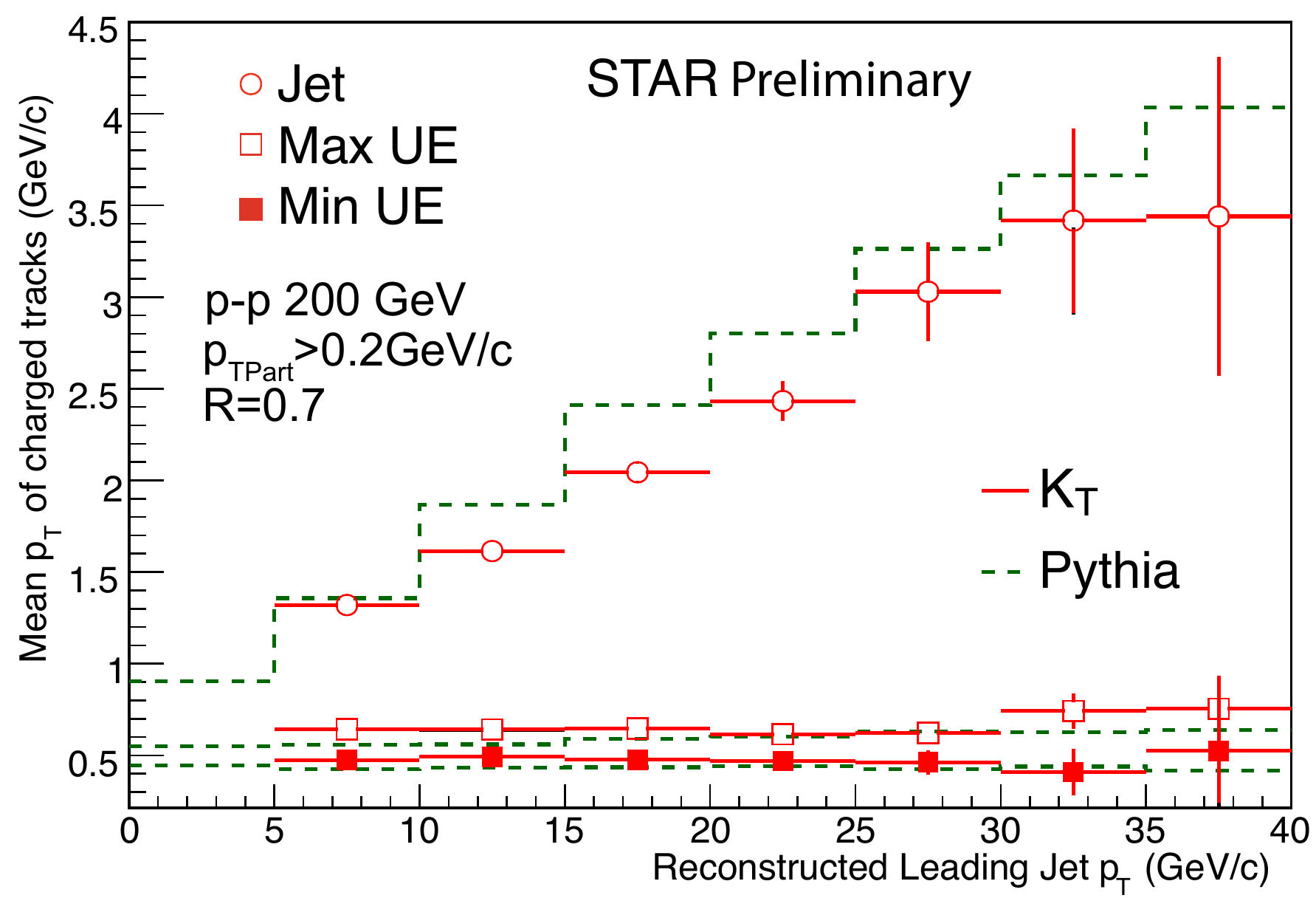}
		\end{center}
	\end{minipage}
	\begin{minipage}{0.46\linewidth}
		\begin{center}
		\includegraphics[width=\linewidth]{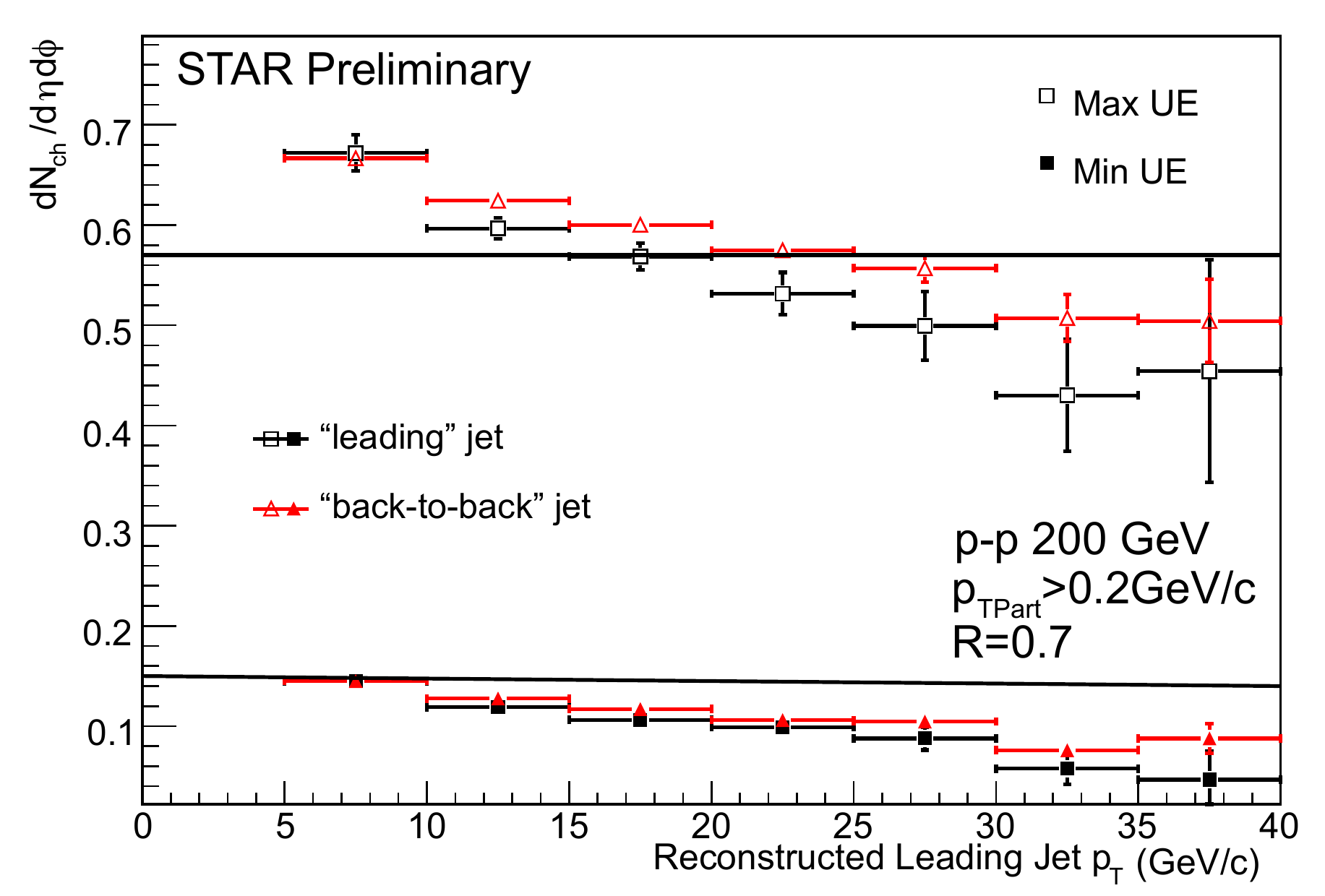}
		\end{center}
	\end{minipage}
	\caption{ Color online: Uncorrected charged particle results  as a function of  reconstructed lead jet \pT, for R=0.7. Left) The mean \pT\ of the jet and the TransMin and TransMax regions of the underlying event  for data and PYTHIA. Right) The  particle density in the TransMin and TransMax regions for both leading jet and di-jet event selections. The solid lines assume the particle density is driven by a Poisson distribution with an average of 0.36. }
			\label{Fig:UE}
\end{figure}

The particles resulting from a p-p collision are not only from hard scatterings.  Initial and final state radiation, soft and semi-hard multiple parton interactions and beam-beam remnants also contribute to the overall multiplicity. The contributions from everything except the hard scattering are termed the underlying event (UE). Pile-up is not part of the definition of the UE.  For this UE study we follow the technique developed by CDF~\cite{Cruz:2005ru}.  Once jets have been reconstructed, each event is divided into four sections defined by their azimuthal  angle with respect to the leading jet axis ($\Delta{\phi}$).  The range within $|\Delta{\phi}|$$<$60$^{0}$ is the lead jet region and that  for $|\Delta{\phi}|$$>$120$^{0}$ is the away jet area. The two  sectors defined using $60^{0}$$<$$\Delta{\phi}$$<$120$^{0}$ and $-120^{0}$$<$$\Delta{\phi}$$<$-60$^{0}$ are the transverse regions. The transverse sector containing the largest charged particle multiplicity  is  called the TransMax region, and the other is  termed the TransMin region. The TransMax region has an enhanced probability of containing contributions from the hard initial and final state radiation components.  The contribution from beam-beam remnants is believed to be negligible since the measurements are performed at mid-rapidity. The mean \pT\ of charge particles in the jet, TransMax, and TransMin regions for ``leading" jet events are shown in left panel of Fig.~\ref{Fig:UE}. Leading jet events are those where at least one jet is found in the STAR acceptance. As expected the $\langle p_{T} \rangle$ of particles in the jet is much higher than in the transverse regions. While the  $\langle p_{T}^{TransMax}\rangle > \langle p_{T}^{TransMin} \rangle$ both are independent of the jet \pT\ within errors.  Detector level PYTHIA predictions show reasonable agreement to the data. To suppress hard initial and final state radiation contributions to the UE a ``back-to-back" sub-set of events is selected which have  two (and only two) found jets  with $p_{T}^{awayjet}/p_{T}^{leadjet}$$>$0.7 and $|\Delta{\phi_{jet}}|$$>$150$^{0}$.  By comparing measurements from the TransMax and TransMin regions in the ``leading" and ``back-to-back" sets we can extract information about the various components of the UE. The charged particle densities in the UE regions are shown in the right hand plot of Figure~\ref{Fig:UE}. As with the  $\langle p_{T} \rangle$, the particle density is largely independent of the leading jet energy. The charged particle densities are the same within errors for the ``leading" and ``back-to-back" datasets, indicating that only very small  amounts of  initial/final state radiation are emitted at large angle at RHIC energies. This is in stark contrast to the ``leading"/``back-to-back" density ratio of $\sim$0.65 recored at 1.96 TeV~\cite{Cruz:2005ru}.  The black lines show the expected densities assuming  that the UE distributions are  Poissonian with an average of 0.36. The similarity of this simple simulation to the data suggests that at RHIC energies the splitting of the measured TransMax and TransMin values  is predominantly due to the statistical sampling. Although not shown in Fig.~\ref{Fig:UE} PYTHIA  shows satisfactory agreement  when Tune A is used for both ``leading" and ``back-to-back" data.

\section{Summary and Conclusions}

In summary, various properties of back-to-back jets from data and simulation have been compared.  Good agreement is observed and indicates that STAR's simulations of jets,  including detector effects,  are well controlled for p-p events at \sqrts =200 GeV. A relative jet energy resolution of   $10-15\%$ has been deduced.
Preliminary jet fragmentation functions have been measured  for both unidentified charged particles and strange hadrons. They will provide a stringent baseline for the measurements underway in Au-Au collisions. PYTHIA, tuned to 1.96 TeV data, shows reasonable agreement to the unidentified particle fragmentation functions suggesting that the energy dependence of the underlying physics is well modeled. However the details, such as strange particle production are poorly described. Finally, studies of the UE  are underway, and show that it is largely independent of the momentum transfer of hard scattering and receives only minor contributions from initial and final state radiation from these hard scatterings.

\section*{Acknowledgements}
The author wishes to acknowledge the contributions of the Bulldog High Performance computing facility of Yale University to this work, and the support of the DoE.



\bibliographystyle{elsarticle-num}
\bibliography{JetCainesHP}



\end{document}